\documentclass[aps,pre,showpacs,amsfonts,amsmath,twocolumn,floatfix,superscriptaddress]{revtex4}
\usepackage[final]{graphicx}
\usepackage{color}
\usepackage{bm}
\usepackage{float}

\begin{document}

\title{Temperature Distribution and Heat Radiation of Patterned Surfaces \\
  at Short Wave Lengths}

\date{\today}

\author{Thorsten Emig}

\affiliation{Massachusetts Institute of Technology, MultiScale Materials Science
for Energy and Environment, Joint MIT-CNRS Laboratory (UMI 3466),
Cambridge, Massachusetts 02139, USA}
\affiliation{Laboratoire de Physique
Th\'eorique et Mod\`eles Statistiques, CNRS UMR 8626, B\^at.~100,
Universit\'e Paris-Saclay, 91405 Orsay cedex, France}

\begin{abstract}
  We analyze the equilibrium spatial distribution of surface
  temperatures of patterned surfaces. The surface is exposed to a
  constant external heat flux and has a fixed internal temperature
  that is coupled to the outside heat fluxes by finite heat
  conductivity across surface. It is assumed that the temperatures are
  sufficiently high so that the thermal wavelength (a few microns at
  room temperature) is short compared to all geometric length scales
  of the surface patterns. Hence the radiosity method can be
  employed. A recursive multiple scattering method is developed that
  enables rapid convergence to equilibrium temperatures.  While the
  temperature distributions show distinct dependence on the detailed
  surface shapes (cuboids and cylinder are studied), we demonstrate
  robust universal relations between the mean and the standard deviation
  of the temperature distributions and quantities that characterize overall
  geometric features of the surface shape.
\end{abstract}


\maketitle

\section{Introduction}

Planck's law describes the intensity of radiation of a black body with
temperature $T$ at a given wavelength
\cite{Planck:1901ul}. Integration over all wavelengths yields the
Stefan-Boltzmann law \cite{Boltzmann:1884wd} for the total power  $P$
emitted by the black body
\begin{equation}
  \label{eq:6}
  P = \sigma A T^4
\end{equation}
where $A$ is the surface area of the body, and
$\sigma=\pi^2 k_B^4/(60\hbar^3 c^2)$. For real materials
Eq.~(\ref{eq:6}) is modified by multiplying $\sigma$ with the
emissivity of the material. However, recently various modifications of
the radiation law due to size and shape of the body have been explored
and new general approaches based on scattering theory have been
developed \cite{kruger2012trace}. In general, the (effective)
emissivity of an object depends on its size and shape due to
self-scattering of the emitted radiation. Recent scattering
approaches, however, assume that the bodies' surface has a spatially
constant temperature.  In general, this is not strictly justified due
to self-absorption of heat emitted by a body with a non-planar surface.

Information about the temperature distribution on patterened objects
and the resulting transport of energy by heat radiation
\cite{Modest:ij} is important to many science and engineering
applications: radiating micro-structured surfaces, transfer in
combustion chambers and heat exchangers, climate phenomena like the
spatial variation of land surface temperatures
\cite{ISI:A1981NM30200003}, solar energy utilization and the design of
sustainable buildings.  Modeling of heat radiation and radiative heat
transfer in large-scale, complex geometries consisting of many shapes,
objects and materials presents enormous challenges due to the
long-range wave nature of electromagnetic radiation.  Most precise
solution requires numerical solution of the electromagnetic wave
equation to obtain the scattering of electromagnetic waves at all
surfaces. However, for large complex geometries, the computing time
and lack of precision of this methods increases
\cite{Howell:1998nr}. Hence, it is desirable to identify universal
scaling laws that can predict how shape and geometry influences
spatial variation of temperatures and heat radiation. This work
attempts to propose a step in this direction by considering surfaces
with various geometric patterns.

We assume that the thermal wavelength $\lambda_T=\hbar c/(k_B T)$ is
short compared to all geometric length scales of the surface patterns.
In this limit, geometric optics can describe heat radiation leading to
the so-called radiosity method that is widely used for heat phenomena
and visual rendering \cite{Gortler:1993:WR:166117.166146}. It assumes
diffuse reflections at the surfaces and hence is an alternate method
to ray tracing. The surface is decomposed into patches that are
coupled via a so-called view factor matrix that measures the fraction
of radiation that travels from one surface patch to another. Similar
methods can be applied to interactive sound propagation in complex
environments (urban or indoor environments such as auditoriums)
\cite{Schissler:2014qf}.

\section{The Model}

We consider a geometrically structured two-dimensional surface that is
decomposed into small surface ``patches'' given by $N$ mutually
joining polygons $P_j$, $j=1,\ldots, N$, defined over a planar base
plane ($xy$-plane). The polygons are oriented so that their surface
normals ${\bf n}_j$ are pointing all into the same half-space, the
``outside'', (say the positive $z$-direction) which contains the
source of the incoming external heat flux.  For simplicity, we assume
further that the polygon surface normals are either normal or parallel
to the base plane.  Each polygon is further characterized by an
emissivity $\epsilon_j$, surface thickness $d_j$, and thermal
conductivity $\kappa_j$. On the ``inside'' (negative $z$-direction) of
the surface a local equilibrium inside temperature $T^\text{int}_j$ is
imposed for each polygon.  We assume that the surface receives a
homogeneous radiant flux $L$ from the outside half-space or
``sky''. The goal is to compute the equilibrium temperatures $T_j$
on the outside surfaces of the polygons assuming that they are
insulated against each other. These temperatures are determined by
equating the internal and external net flux densities for each
polygon. The internal net flux is obtained from the stationary heat
conduction equation $q^\text{int}_j=-\kappa \partial_n T_j$ integrated
across the surface thickness $d_j$ yielding
$q^\text{int}_j=(T_j-T^\text{int}_j)\kappa_j/d_j$. The external net
flux $q^\text{ext}_j$ is obtained as the sum of the incoming fluxes
from the sky ($L$) and those scattered from all other visible polygons
and the heat flux $\sigma \epsilon_j T_j^4$ radiated by the surface
$j$ where $\sigma$ is the Stefan-Boltzmann constant.

For the simple case of a single planar surface ($j=N=1$), the
condition $q^\text{ext}_1=q^\text{int}_1$ yields
\begin{equation}
  \label{eq:1}
  (T_\text{flat}-T^\text{int})\frac{\kappa}{d} = \epsilon (L-\sigma
  T_\text{flat}^4) \, ,
\end{equation}
which determines the outside surface temperature $T_\text{flat}$ of the flat surface
as function of known parameters. 

For a general structured surface one has to consider multiple
reflections between surface patches that contribute to the net external
fluxes. To describe this effect, it is assumed that the surface patches
are gray diffusive emitters, i.e., the emissivity is frequency
independent and the radiation density is constant across the 
surface patches and emitted independent of direction. We expect this to be a
good approximation for thermal wavelengths that are small compared to
the geometric structure of the surface and hence the size of the
patches. Then we can apply to radiosity concept to obtain the
external fluxes $q^\text{ext}_j$\cite{Modest:ij}. For a given
surface patch $j$, the outgoing radiant flux is given by the sum
of emitted thermal radiation and the reflected incoming radiation,
\begin{equation}
  \label{eq:2}
  J_j = \sigma \epsilon_j T_j^4 + (1-\epsilon_j) E_j 
\end{equation}
where we used that the reflectivity equals $1-\alpha_j$ for an opaque
surface where $\alpha_j=\epsilon_j$ is the absorptivity.  How much
energy two surface patches exchange via heat transfer depends
on their size, distance and relative orientation which are encoded in
the so called view factor $F_{ij}$ between patches $i$ and
$j$. $F_{ij}$ is a purely geometric quantity and does not depend on
the wavelength due to the above assumption of diffusive surfaces. 
It is defined by the surface integrals
\begin{equation}
  \label{eq:3}
  F_{ij} = \int_{A_i} \int_{A_j} \frac{\cos \theta_i \cos
    \theta_j}{\pi A_i |\bm{ r}_{ij}|^2} dA_i dA_j
\end{equation}
where $\theta_{i}$ is the angle between the surface patch's normal vector
$\bm{ n}_{i}$ and the distance vector $\bm{ r}_{ij}$ which connects
a point on patch $i$ to a point on patch $j$, and $A_{i}$ is
the surface area of patch $i$ . The view factor matrix obeys the
important reciprocity relation $A_j F_{ji}=A_i F_{ij}$ and additivity
rule $\sum_j F_{ij} = 1$. With this geometric quantity, the radiative
flux received by surface patch
$j$ from all other surface patches can be expressed as
$E_j = \sum_{i} F_{ji} J_i$, and one can solve Eq.~(\ref{eq:2}) for
the vector of outgoing fluxes, yielding
\begin{equation}
  \label{eq:4}
  \bm{J} =  \left[ \bm{1} - ( \bm{1} - \bm{\epsilon})
      \bm{F}\right]^{-1} \bm{J}_0 \, ,
\end{equation}
where we combined the fluxes $J_j$ from all patches into a vector $\bm{J}$
and the radiation $\sigma \epsilon_j T_j^4$ into a vector ${\bf J}_0$ to
use a matrix notation. Here $\bm{1}$ is the identity matrix and
$\bm{\epsilon}$ the diagonal matrix with elements $\epsilon_j$.
To compute the surface temperaturs $T_j$ we need to compute the net heat transfer
to surface patch $j$ which is given by the incident radiation $E_j$ minus
the outgoing flux $J_j$, leading to the net flux $q_j^\text{ext} = \sum_i F_{ji}
J_i - J_j$. In vector notation this net flux becomes
\begin{equation}
  \label{eq:5}
  \bm{q}^\text{ext} = ( \bm{F} - \bm{1} ) \left[ \bm{1} - ( \bm{1} - \bm{\epsilon})
      \bm{F}\right]^{-1} \bm{J}_0 \, .
\end{equation}
In the stationary state, the surface patch temperatures are then
determined by the condition that the net external flux equals the net
internal flux, $\bm{q}^\text{ext} = \bm{q}^\text{int}$ where
$\bm{q}^\text{int}$ defines the vector with elements
$(T_j-T_j^\text{int})\kappa_j/d_j$ due to heat conduction across the
surface (see above). This condition uniquely fixes the temperatures
$T_j$ when all other parameters including the external (``sky'') flux
$L$ are known. In the following, technically we include the ``sky'' as
an additional surface so that we have now $N+1$ surface patches. The
corresponding additional matrix elements for the view factor matrix
$\bm{F}$ follow from reciprocity and additivity rules, and we include the
downward radiation $L$ as the $(N+1)^\text{th}$ component in
$\bm{J}_0$.

Knowing the surface temperatures, a number of interesting observables
can be obtained. An {\it effective emissivity} of the total surface
can be defined as the ratio $\epsilon_\text{eff} = Q/Q_\text{bb}$
where
$Q=[\bm{F}\left[ \bm{1} - ( \bm{1} - \bm{\epsilon}) \bm{F}\right]^{-1}
\bm{J}_0]_\text{j="sky"} $
is the net flux towards the ``sky'' and
$Q_\text{bb}=[\bm{F J}_\text{bb}]_\text{j="sky"}$ is again a net flux to
the ``sky'' but assuming that all surface patches radiate as ideal
black bodies, corresponding to
$\bm{J}_\text{bb}=\sigma[T_1,\ldots,T_N,0]$. An {\it effective
  temperature} $T_\text{eff}$, as observed from the ``sky'', can now
be defined as were all surfaces black bodies at their local
temperature, so that $\sigma T_\text{eff}^4 = Q_\text{bb}$ and
$Q=\sigma \epsilon_\text{eff} T_\text{eff}^4$. We also define the difference 
$\Delta T = T_\text{eff}- T_\text{flat}$.

\section{Numerical Implementation}

\begin{figure}[H]
\includegraphics[width=1\linewidth]{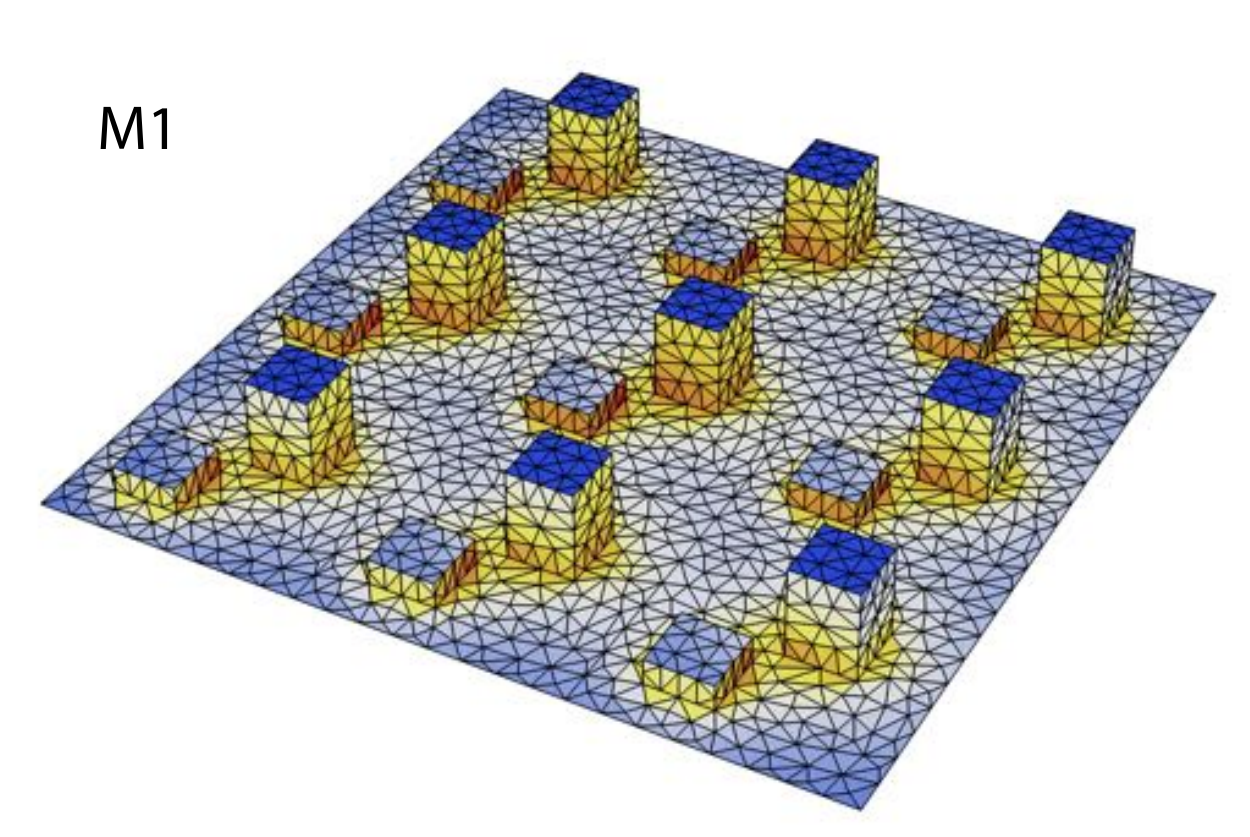}
\includegraphics[width=1\linewidth]{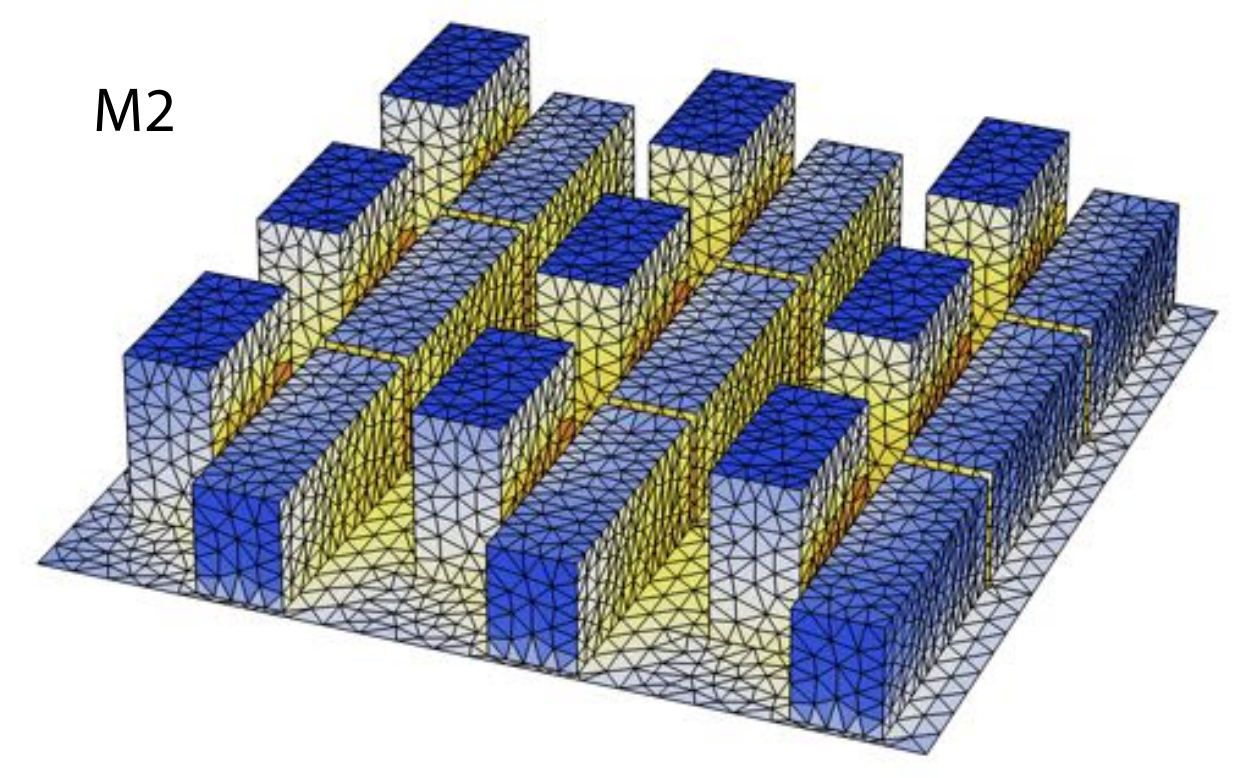}
\includegraphics[width=1\linewidth]{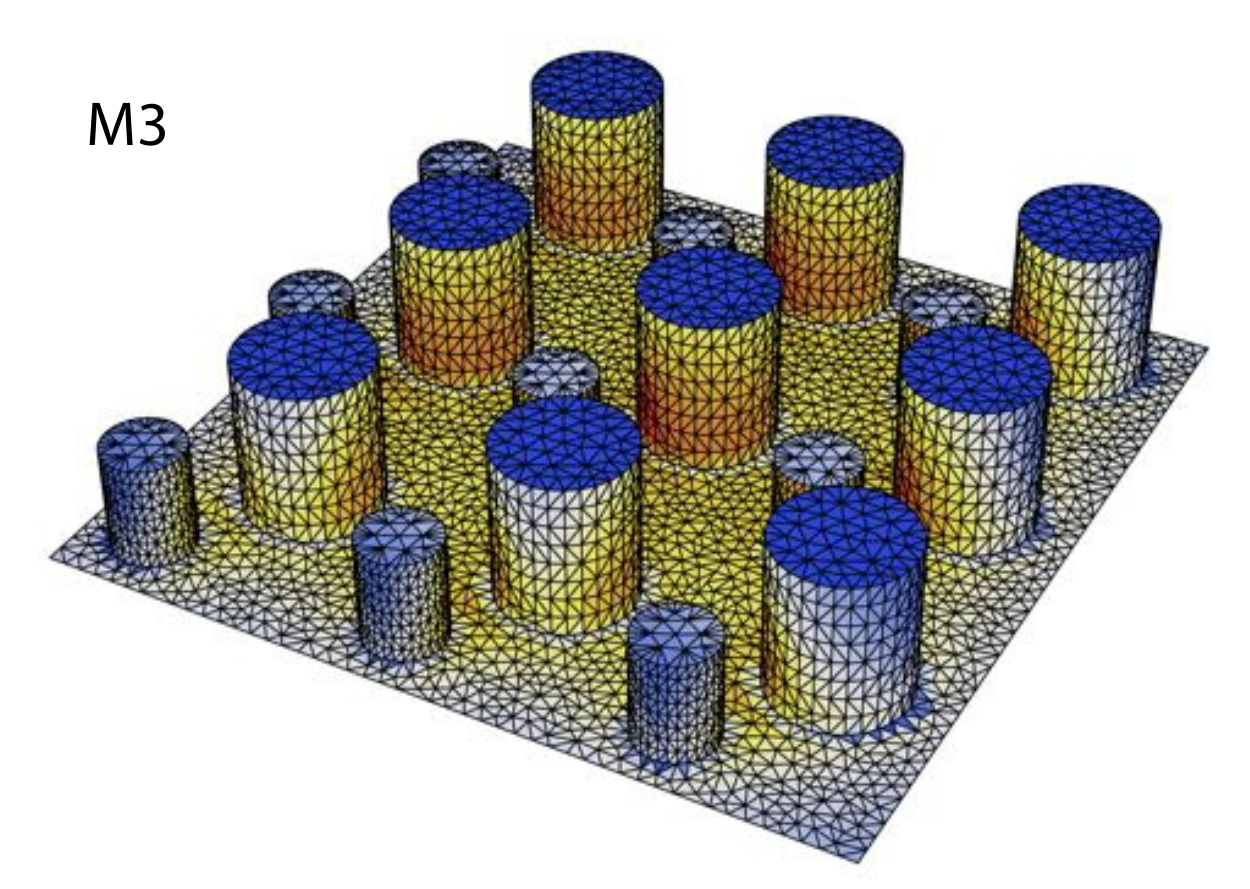}
\caption{\label{fig:1} Surface patch temperature distribution for models M1,
M2, and M3. Colors represent temperature changes from minimum (blue)
to maximum (red) temperature. For values see histograms in
Figs.~\ref{fig:2} to \ref{fig:4} and Tab.~\ref{tab:models}.}  
\end{figure}

  \begin{table*}[t]
    \centering
    \begin{tabular}{|r|rrrrr|r|rrrr|rrrr|rrr|}
\hline
    model  & $A_g$ & $A_v$ & $\bar F_\text{all $\!\to\!$ sky}$ & $\bar F_\text{b $\!\to\!$ sky}$ & patches & $T_\text{flat}$ & $\bar T$ &
                                                                    $\bar T_v$
      & $\bar T_t$& $\bar T_b$ & $\sigma$ &  $\sigma_v$ &  $\sigma_t$ &
                                                                    $\sigma_b$
      & $T_\text{eff}$ & $\epsilon_\text{eff}$ & $\Delta T$ \\
\hline
     M1 & 50 & 180 & 0.7175 & 0.8037 & 4140 &&&&&&&&&&&& \\
     $\epsilon=0.5$ & & & & & & 285.49 & 286.45 & 287.20 & 285.61 &
                                                                 286.20
                               & 0.64 & 0.34 & 0.13 & 0.39 & 292.74 &
                                                                      0.5631
                                               & 7.26 \\
     $\epsilon=0.9$ & & & & & & 282.31 & 284.13 & 285.55 & 282.54 & 283.65
                                                                 
                               & 1.20& 0.60& 0.24 & 0.72 & 290.12&
                                                                   0.9193& 7.81 \\
\hline
     M2 & 186& 832 & 0.3441 & 0.3171 & 9018 &&&&&&&&&&&& \\
     $\epsilon=0.5$ & &&&&&285.49&289.04&289.77&285.79&289.29&1.83&1.30&0.28&0.93&315.76&0.6113&30.27 \\
     $\epsilon=0.9$ & &&&&&282.31&288.04&289.20&282.85&288.42&2.79&1.82&0.49&1.26&313.32&0.9293&31.00 \\
\hline
     M3 & 106.81 & 527.79 & 0.4780 &  0.5751 & 14211 &&&&&&&&&&&& \\
     $\epsilon=0.5$ & &&&&&285.49&287.63&288.07&285.59&287.30&0.83&0.48&0.16&0.40&315.13&0.5964&29.64 \\
     $\epsilon=0.9$ & &&&&&282.31&286.12&286.91&282.51&285.52&1.47&0.82&0.29&0.69&312.62&0.9277&30.31 \\
\hline
    \end{tabular}
    \caption{Geometric parameters and surface temperature characteristics for the three surface models. All
      temperatures and their standard deviations are given in Kelvin.}
    \label{tab:models}
  \end{table*}

The numerical implementation of the model described above follows
these steps:
\begin{enumerate}
\item The surface is decomposed into oriented patches which is done here by
  triangularization so that the entire surface is composed of planar
  triangular surface elements, see Fig.~\ref{fig:1} with their surface
  normal vector pointing to the ``outside'' of the surface, i.e.,
  pointing towards the ``sky''. For later analysis, these elements are
  grouped into three different classes: horizontal ``base'' patches
  (b) that are located within the base plane $z=0$, horizontal ``top''
  patches (t) that are located above the base plane and ``vertical''
  patches (v) that are perpendicular the base plane and connect the
  patches in class b and t.

\item Determine for all pairs of patches if the view between them is
  blocked by other patches. This is done by testing for potential
  intersections of the ray connecting the two centroids of a pair of patches
  and all other surface patches. It is sufficient to perform this
  visibility test for pairs of patches of the type $(v, b)$,  $(v,t)$
  and $(v,v)$ where the first (second) letter denotes the class of the
  first (second) patch. For all these combinations potential blocking
  patches must be in class $v$. 

\item If the view between a pair $(i,j)$ of patches is not blocked and
  the first patch can ``see'' the outside of the second, the view
  factor $F_{ij}$ is computed, using the exact closed form expression
  described in \cite{Schroder:1993:FFT:166117.166138}. This is done
  for all patch class combinations $(v, b)$, $(v,t)$ and $(v,v)$ with
  the restriction $i<j$ for $(v,v)$ since the view factors for $i>j$
  follow from reciprocity.

\item Construct the total view factor matrix $\bm{F}$ for all patches
  of classes $v$, $b$ and $t$ and the single enclosing surface
  describing the ``sky''. This is done by using reciprocity to obtain
  the matrix elements for the patch class combinations $(b,v)$ and
  $(t,v)$. The patches of classes $b$ and $t$ cannot see each other so
  that the view factor submatrix for these classes vanishes. To obtain
  the view factor for the transfer from a surface patch $i$ towards
  the ``sky'' we use the sum rule $\sum_j F_{ij}=1$, i.e., $F_{i
    \,\text{"sky"}}= 1- \sum_{j \in \{b,t,v\}} F_{ij}$. The view
  factor for the transfer from the ``sky'' to a patch $i$ follows from
  reciprocity as $F_{\text{"sky"}\, i}= \frac{A_i}{A} F_{i
    \,\text{"sky"}}$ where $A$ is the total area of the surface. 

\item The inverse matrix of Eq.~(\ref{eq:5}) can be computed as a
  truncated geometric series since the emissivities are sufficiently
  close to unity and the view factors $F_{ij}<1$ with most of them in
  fact much smaller then unity. Hence the inverse kernel is given by
  $\bm{K}^{-1} \equiv \left[ \bm{1} - ( \bm{1} - \bm{\epsilon})
    \bm{F}\right]^{-1}=\sum_{n=0}^{n_c} \bm{M}^n$
  with $\bm{M}=( \bm{1} - \bm{\epsilon}) \bm{F}$. We find that $n_c=6$
  is sufficiently accurate approximation for the parameters used
  below.

\item Finally, we compute the surface patch temperatures $T_j$ by an
  iterative solution of the equilibrium condition
  $\bm{q}^\text{ext} = \bm{q}^\text{int}$ [see Eq.~(\ref{eq:5})] for
  given surface emissivities $\epsilon_j$, downward radiation $L$,
  interior temperatures $T^\text{int}_j$ and effective thermal
  conductivities $\kappa_j/d_j$. The iteration steps are as follows:

\begin{itemize}

\item[(i)] Choose initial patch temperatures $T_j^{(\nu=0)}$. 

\item[(ii)] Compute the external flux
  $\bm{q}^{\text{ext}\, (\nu=0)} = (\bm{F} - \bm{1}) \bm{K}^{-1}
  \bm{J}^{(\nu=0)}_0$
  with the $N+1$ dimensional initial vector
  $\bm{J}^{(\nu=0)}_0=[L,\sigma \epsilon_1 {T_1^{(\nu=0)}}^4, \ldots ,
  \sigma \epsilon_N {T_N^{(\nu=0)}}^4]$.

\item[(iii)] Compute the updated patch temperatures $T_j^{(\nu=1)}$
  from the equation $q_j^{\text{ext}\, (\nu=0)} = ( T_j^{(\nu=1)} -
  T^\text{int}_j ) \kappa_j/d_j$ for $j=1,\ldots, N$.

\item[(iv)] Continue with step (i) to start the next iteration step,
  i.e.,
  $\bm{q}^{\text{ext}\, (\nu=1)} = (\bm{F} - \bm{1}) \bm{K}^{-1}
  \bm{J}^{(\nu=1)}_0$
  with the vector
  $\bm{J}^{(\nu=1)}_0=\{L,\sigma \epsilon_1 [(T_1^{(\nu=1)}+T_1^{(\nu=1)})/2]^4, \ldots ,
  \sigma \epsilon_N [(T_N^{(\nu=1)}+T_N^{(\nu=1)})/2]^4\}$.
\end{itemize}

In (iv) and all following iteration steps it is useful to use the
average of the last two iterations for the patch temperatures, as
indicated here, to obtain rapid convergence. Typically, for the models
and parameters used below, after about 20
iterations a stable solution for the patch temperatures had been
reached (within a relative accuracy of $10^{-4}$). 

\end{enumerate}

\section{Results}

In order to study the influence of the density and shape of surface
patterns on the temperature distribution, we have considered three
different surface structures that are all periodic in both spatial
directions, see Fig.~\ref{fig:1}. For all surfaces, the dimension of a
unit cell given by $L_x\times L_y= 20 \times 20$ (in arbitrary units).
It is assumed that all spatial dimensions, however, are large compared
to the thermal wavelengths $\lambda_T=\hbar c/(k_B T)$ of the
surface temperatures which is in the range of a few microns for the
temperatures considered below.  The downward radiant flux from the
``sky'' is set to $L=300$W per unit surface area, the interior surface
temperatures are all set to the temperature
$T_j^\text{int}\equiv T^\text{int}=293.15^\circ$K, and all surface
thicknesses $d_j$ and thermal conductivities $\kappa_j$ are chosen
such that ratio $\kappa_j/d_j=5.0$W/K per unit surface area.  We
consider two different homogenous emissivities across all surface
patches which are $\epsilon=0.5$ and $\epsilon=0.9$.

The resulting surface temperature distributions for the three
different geometric pattern are shown in Fig.~\ref{fig:1}. The
geometric characteristics of the models are as follows: each model is
composed of $9$ unit cells. Model M1's unit cell consists of two
rectangular cuboids with dimensions $5 \times 5 \times 2$ and
$5 \times 5 \times 7$, respectively. Model M2's unit cell is composed
of two rectangular cuboids with dimensions $6 \times 19 \times 8$ and
$6 \times 12 \times 12$, respectively. Finally, the unit cell of model
M3 is composed of two cylinders of radii $r_j$ with dimensions
$r_1=3 \times 8$ and $r_2=5 \times 12$, respectively.  The
corresponding area $A_g$ (per unit cell) of the base plane that is
covered by these elements (cuboids, cylinders) and the area $A_v$ (per
unit cell) of their vertical surfaces are summarized in
Tab.~\ref{tab:models}. In that table the total number of surface
patches is also indicated. As we shall see below, other important
geometric quantities are certain averaged view factors: the average
``sky'' view $\bar F_\text{all $\to$ sky}=\sum_{j\in \{b,t,v\}} F_{j\,
  \text{"sky"}}/N$
from all surface patches, and the average ``sky'' view
$\bar F_\text{b $\to$ sky}=\sum_{j\in \{b\}} F_{j\, \text{"sky"}}/N_b$
from patches of the base plane only, where $N_b$ is the number of base
plane patches. These averages were restricted to the central unit cell
to avoid boundary effects and they are also given in
Tab.~\ref{tab:models}.

Next we analyze the results for the temperature distributions as they
follow from the numerical approach outlined above. As can be seen from
Fig.~\ref{fig:1}, the coldest patches are those on the top of the
structures (class t). Since the top patches of the highest structures
do not interact with any other patches, their temperature equals the
temperature $T_\text{flat}$ of a planar surface which sets hence the
minimum value for the temperature distribution. Highest temperatures
are observed on the vertical surface patches with an increase in
temperature from the top to the bottom. This pattern results from a
decreased view of open space (``sky'') for vertical patches and
reflections from the base patches close to the bottom of the elevated
structures. The base patches' temperature decays away from the
structures which is clearly visible for the low structures of model
M1. The non-central unit cells show colder surface patches towards the
edges of the surface due to their proximity to the boundaries which
enables an increased emission of heat.

\begin{figure}[H]
\includegraphics[width=.9\linewidth]{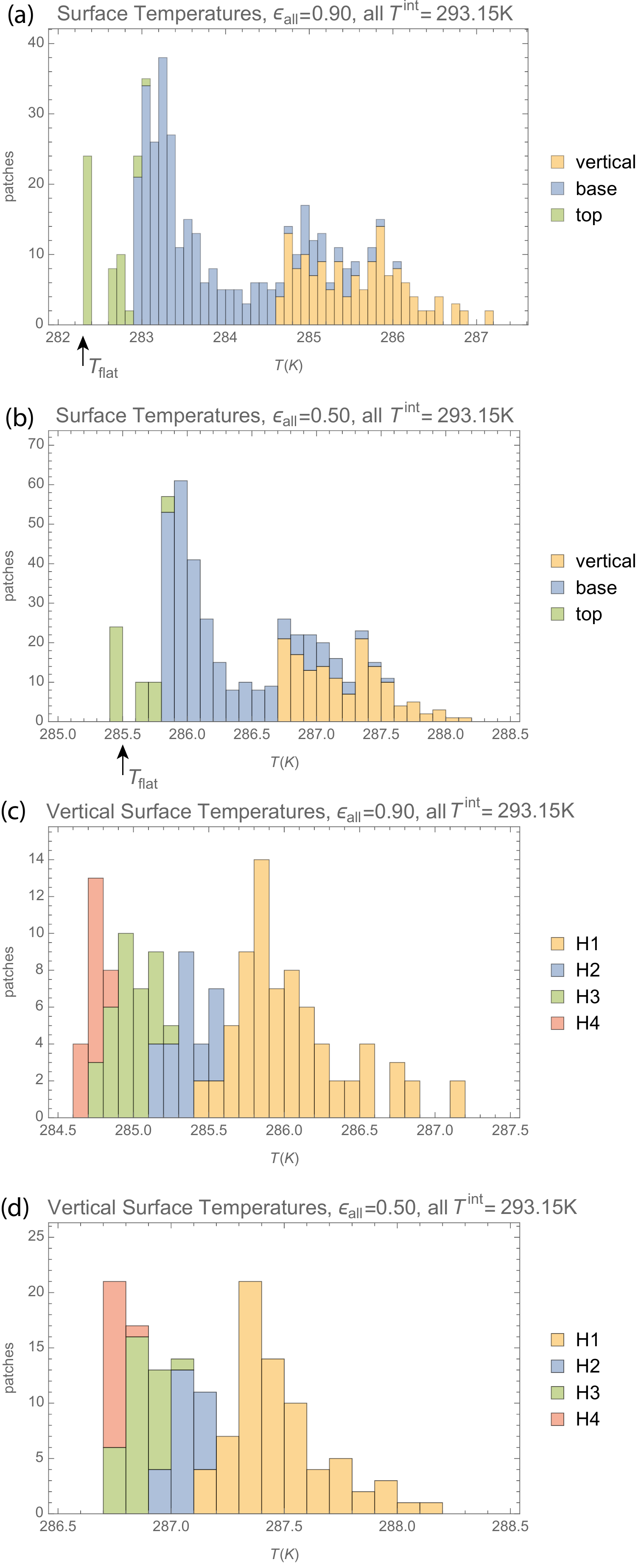}
\caption{\label{fig:2} Histograms for surface patch temperatures of
  the central unit cell of 
  model M1: (a) temperatures for the three different patch classes
  vertical ($v$), base ($b$), and top ($t$) for emissivity
  $\epsilon=0.9$, (b) same as (a) for emissivity $\epsilon=0.5$, (c)
  temperatures for vertical ($v$) patches for emissivity
  $\epsilon=0.9$, grouped into four different equidistance height
  classes $H1$ to $H4$ according to their height over the base plane,
  and (d) same as (c) for emissivity $\epsilon=0.5$.}
\end{figure}

\begin{figure}[H]
\includegraphics[width=.9\linewidth]{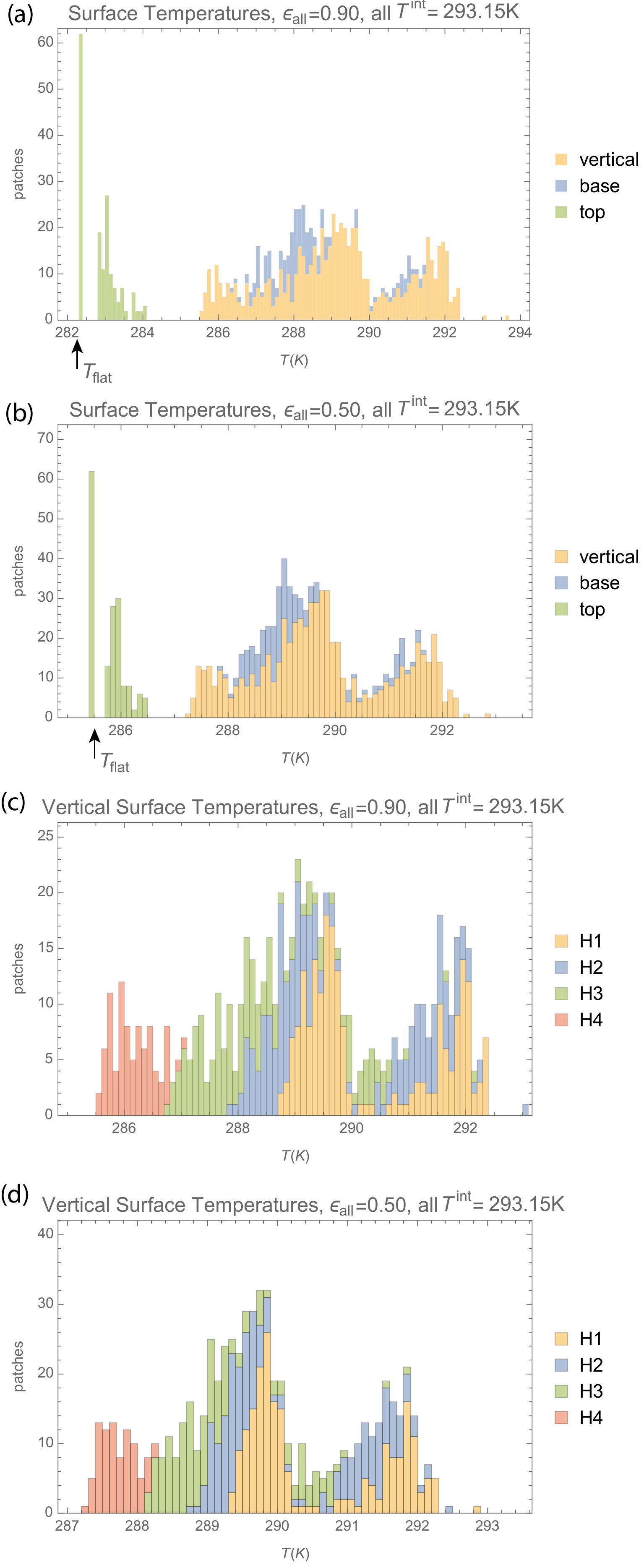}
\caption{\label{fig:3} Histograms for surface patch temperatures as in
  Fig.~\ref{fig:2} for model M2.} 
\end{figure}

\begin{figure}[h]
\includegraphics[width=.9\linewidth]{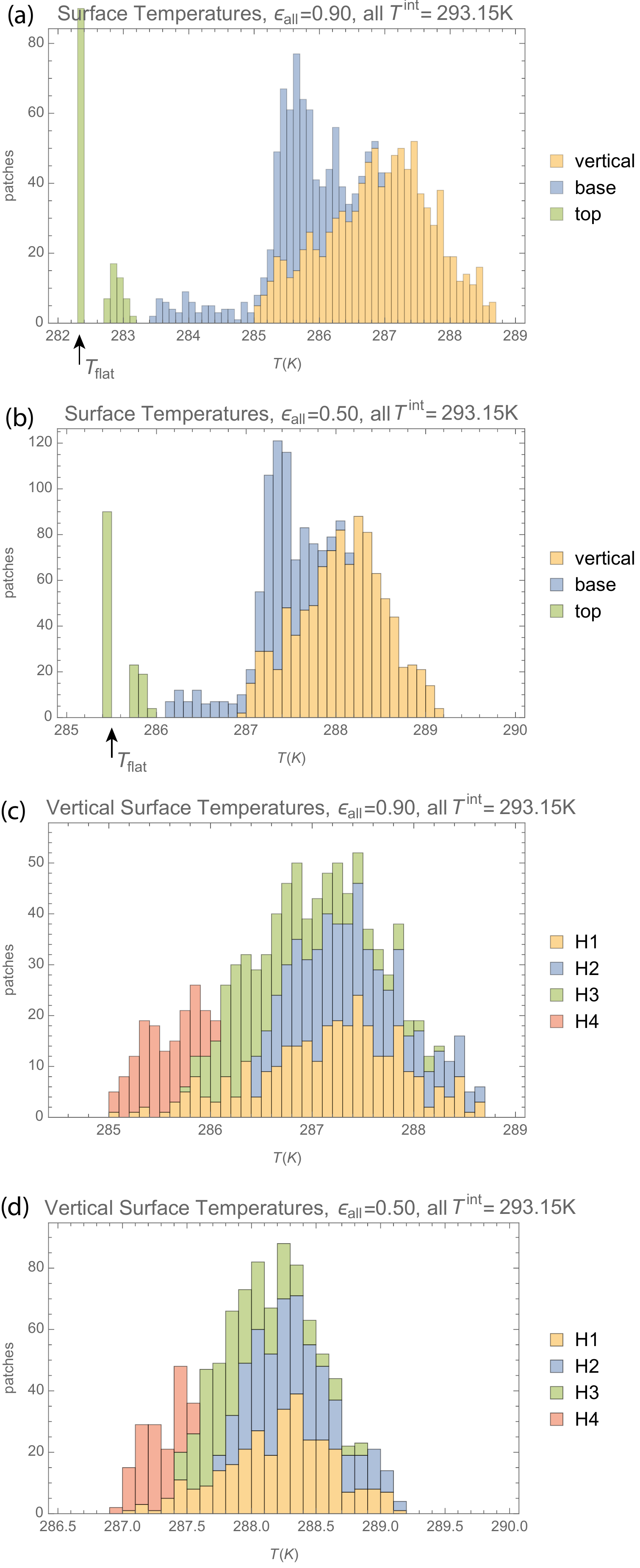}
\caption{\label{fig:4} Histograms for surface patch temperatures as in
  Fig.~\ref{fig:2} for model M3.} 
\end{figure}

Figures \ref{fig:2} -- \ref{fig:4} show histograms for the surface
temperature distributions of the three models, indicating the number
of patches at a given temperature. Different colors label the three
different classes of surface patches: vertical, base, and top patches.
To reduce boundary effects, the histograms show the distribution of
the center unit cell. For all models, panels (a) and (b) show the
entire distribution for $\epsilon=0.5$ and $\epsilon=0.9$,
respectively. Panels (c) and (d) show the distributions for the
vertical patches only, again for $\epsilon=0.5$ and $\epsilon=0.9$,
respectively, with different colors labeling now equidistant height
intervals over the base plane in which the patches are located.
A general feature of all models is that the surface temperatures
increase from top patches over base patches to vertical patches. It is
interesting to note that only for model M1 there is a clear separation
of base and vertical temperature ranges whereas for M2 and M3 the base
temperatures fall into the mid or lower range of vertical
temperatures. Another interesting observation is that the vertical
temperature distribution has a single peak for models M1 and M3,
particularly in the latter, and a two-peak structure for model M2. We
interpret this as a consequence of the proximity of two cuboids of
different height. This view if supported by the variation of the
distribution of vertical temperatures with height, see
Fig.~\ref{fig:3}(c) and (d): Only the peak at smaller temperatures
contains patches of the largest height class H4, and hence must
represent mainly the taller cuboid. In generel, models M1 and M2 
display little overlap between the temperatures corresponding to
different height intervals while model M3 shows less separated
temperature ranges for the height intervals. This is presumably
related to the continuous range of vertical surface patch orientations
for cylinders as compared to cuboids.

Table~\ref{tab:models} summarizes various characteristics of the
temperature distributions. In addition to the quantities
$T_\text{eff}$, $\Delta T$ and $\epsilon_\text{eff}$ defined above,
the mean temperature $\bar T$ of the full distribution and the mean
temperatures $\bar T_j$ of the patch classes $j=v, t, b$ are shown.
The measure the temperature variations across different surface areas,
we have also computed the standard deviation $\sigma$ for the full
distribution and the standard deviations $\sigma_j$ for the different
patch classes. Generally, a surface profile with deeper ``canyons''
leads a trapping of radiation and hence a larger $T_\text{eff}$ which
measures shape effects. Similarly, the effective emissivities
$\epsilon_\text{eff}$ show a larger increase for profiles with narrow
``canyons'' since they render the surface more black due to the
trapping of radiation. A surface with a lower bare emissivity
($\epsilon=0.5$) has a larger shape induced increase in emissivity as
an already highly emissive surface ($\epsilon=0.9$).


An important problem is the identification of geometric parameters
that characterize relevant features of the surface shape and show
a universal relation to certain moments of the surface temperature
distributions. Universal means here that the relation, instead of depending
on particular details of the surface structure, relates to simple
overall features of the surface shape. Potential candidates for such
geometric parameters are listed in Tab.~\ref{tab:models}: The surface
areas $A_g$, $A_v$, and the averaged view factors $\bar F_\text{all $\to$
  sky}$, $\bar F_\text{b $\to$ sky}$. 

\begin{figure}[H]
\includegraphics[width=.9\linewidth]{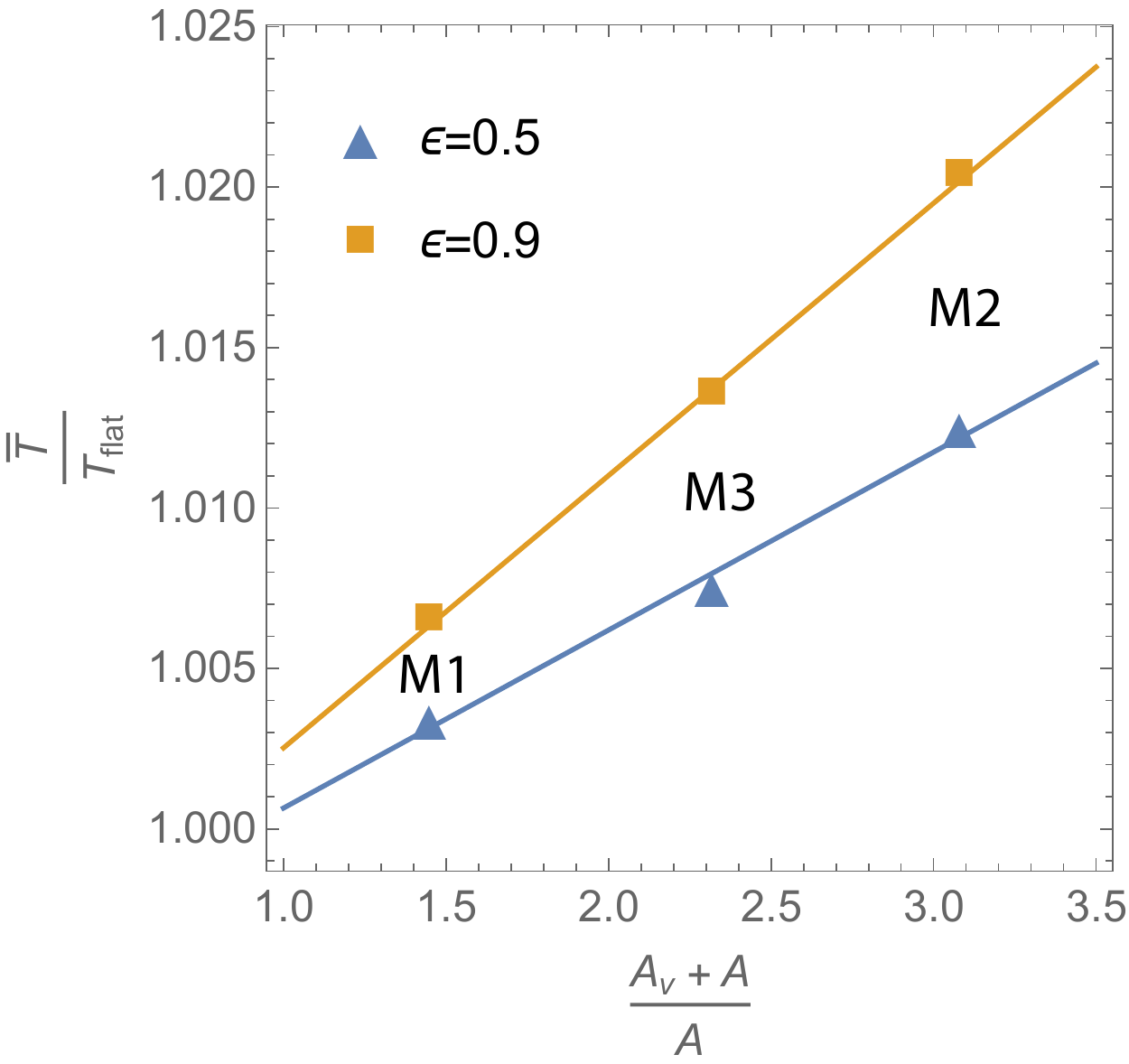}
\caption{\label{fig:5} Mean surface temperature (rescaled by the flat
  surface temperature) as function of the relative increase
  $(A_v+A)/A$ in surface
  area $A=L_x L_y$ due to vertical surface patches.} 
\end{figure}

According to the Stefan-Boltzmann radiation law, the total radiative
power emitted by an ideal black body is proportional to its surface
area. For non-ideal bodies, the radiative power is reduced by an
effective emissivity that depends in general on material, size and
shape of the body. Postulating that multiple reflections of heat
radiation is of sub-leading order for the surface models considered
here, one can expect that the shape induced increase in mean surface
temperature $\bar T$ is proportional to the increase in surface area
due to the surface pattern. Fig.~\ref{fig:5} shows the dependence of
$\bar T$ on the relative increase in surface area (due to vertical
patches of total area $A_v$). Indeed, the data are well described by a
linear scaling, demonstrating that the detailed shape of surface
structures is unimportant for the mean temperature.

\begin{figure}[H]
\includegraphics[width=.9\linewidth]{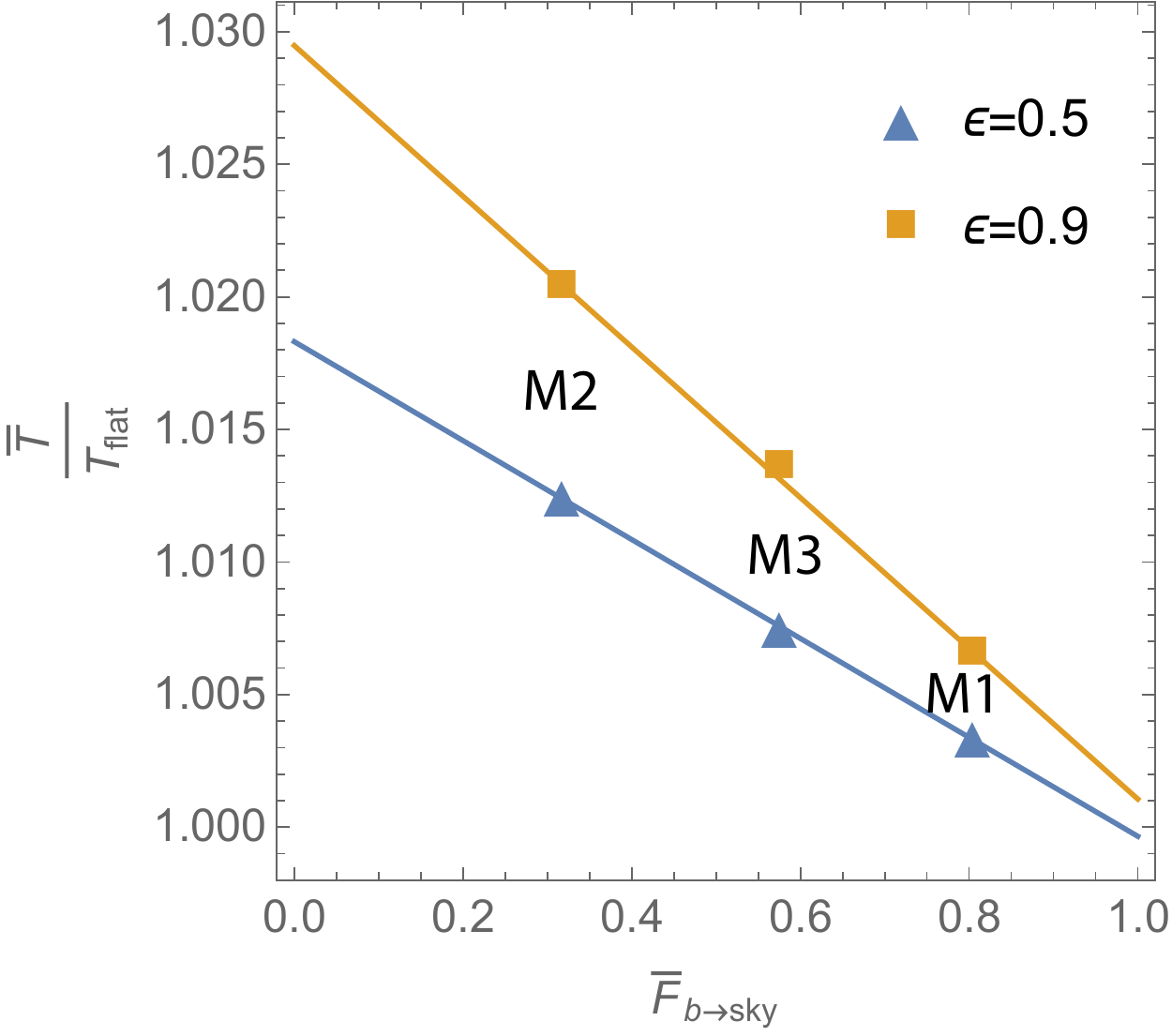}
\caption{\label{fig:6} Mean surface temperature (rescaled by the flat
  surface temperature) as function of the mean view factor $\bar
  F_\text{b $\to$ sky}$ from base
  surface patches towards the ``sky''.} 
\end{figure}

Another geometric quantity that is more sensitive to shape than the
overall increase in surface area is the averaged open (``sky'') view
$\bar F_\text{b $\to$ sky}$ from the base plane patches. For a planar
surface with $\bar T = T_\text{flat}$, the view is unobstructed and
hence $\bar F_\text{b $\to$ sky}=1$. Any surface structure reduces
$\bar F_\text{b $\to$ sky}$ and in fact it has been observed
experimentally in the context of urban climate that mean air and
building surface temperatures tend to increase linearly with a
decrease of the so-called sky-view. To probe this relation
quantitatively, we show in Fig.~\ref{fig:6} the mean surface
temperature as function of the mean open view factor
$\bar F_\text{b $\to$ sky}$. Our data for $\bar T$ show a clear linear
decrease with increasing mean ``sky'' view, with a universal slope
that is independent of the particular surface patterns. The slope,
however, does depend on the emissivity. The total view factor
$\bar F_\text{all $\to$ sky}$, averaged over all surface patches (see
Tab.~\ref{tab:models}) does not show a universal linear relation
across all models.

\begin{figure}[H]
\includegraphics[width=.9\linewidth]{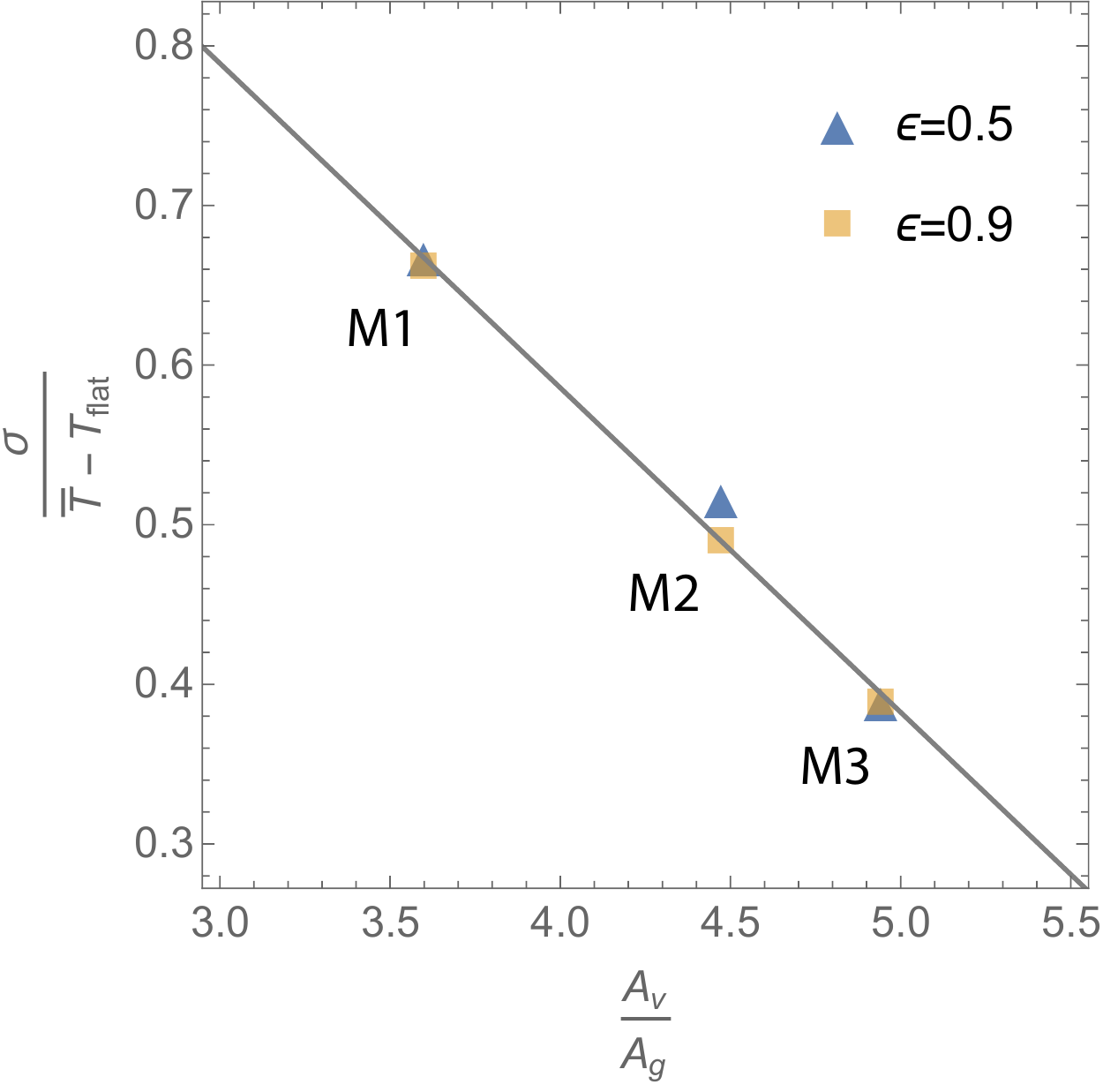}
\caption{\label{fig:7} Standard deviation $\sigma$ of the surface
  temperature distribution, rescaled by the difference
  $\bar T - T_\text{flat}$, as function of the ratio of vertical
  surface area $A_v$ and surface area $A_g$ covered by patterns
  (cuboids, cylinders). Data collapse is observed for different emissivities.}
\end{figure}

Fig.~\ref{fig:1} shows that the temperature distributions have strong
spatial variations.  Hence, it is interesting to identify the key
geometric parameters that determine the statistical moments of the
temperature distributions. We have computed the standard deviation
$\sigma$ of the total distribution which is shown in Fig.~\ref{fig:7},
rescaled by the temperature difference $\bar T - T_\text{flat}$.  The
value of $\sigma$ increases with the emissivity $\epsilon$ which sets
the scale for the typical surface temperatures (which are of course
also dependent on the heat flux from the interior side of the surface,
characterized by the temperature $T^\text{int}$ and heat conductivity
of the surface patches.) However, after the rescaling by
$\bar T - T_\text{flat}$, we observe a convincing collapse of the data
for different $\epsilon$ (see Fig. ~\ref{fig:7}). Interestingly, the
shape dependence of $\sigma/(\bar T - T_\text{flat})$ is controlled by
the ratio of vertical surface area $A_v$ and base surface area $A_g$
covered by elevated structures. This ratio measures the aspect ratio
of height and width of the surface structures, and it shows a linear
relation to $\sigma/(\bar T - T_\text{flat})$. We interpret this
observation as follows: by how much the temperature actually varies
within the typical range between the minimum $T_\text{flat}$ and the
mean $\bar T$ is controlled by the homogeneity of the heat flux
impinging on the surface patches. Tall and thin, antenna like
structures (like the cylinders of model M3) produce a more homogeneous
heat flux (due to their increased view factors) and hence less
temperature variation. This can be observed clearly from the
temperature distribution on the base plane patches in Fig.~\ref{fig:1}
which shows least variation for model M3.

\section{Conclusions}

We have analyzed the influence of geometric surface patterns and
emissivity on the surface temperature distribution, assuming a
homogeneous internal temperature and external radiative flux.  The
surface geometry is assumed to vary on scales large compared to the
thermal wavelengths, i.e., the temperatures have to be sufficiently
large.  The details of the temperature distributions show a rich
structure that is dependent on the detailed surface shape. However, we
could identify parameters that measure relevant overall geometric
features which obey universal relations to the mean and standard
deviation of the surface temperature distributions. It would be
interesting to probe more geometries and a larger range of parameters
to determine the range of validity of these relations.
Also, our study should be extended to non-periodic patterns, and
random surface profiles. There are a number of interesting
conceptional extensions of the approach presented here. For lower
temperatures, or shorter scale surface patterns, diffraction effects
should be added to the radiosity approach. For highly reflective
materials, specular reflections are expected to be important and hence
should be included in the iteraction (view) matrix. Surface geometry
is also expected to modify convective heat transfer which influences
surface temperatures. There is a plethora of possible applications of 
our results ranging from heat transfer between structured surfaces to
the study of climate phenomena.
\\

\begin{acknowledgments}
  Fruitful discussions with M.~Ghandehari are acknowledged.
  The author acknowledges support by the Concrete Sustainability Hub
  at Massachusetts Institute of Technology with
  sponsorship provided by the Portland Cement Association (PCA) and
  the Ready Mixed Concrete (RMC) Research and Education Foundation. 
\end{acknowledgments}

\end{document}